\documentclass[
    ,final            
  ]
  {aipproc}

\layoutstyle{6x9}

\begin{document}

\title{Gamma Ray Bursts as cosmological tools}


\author{G. Ghirlanda}{
  address={Osservatorio Astronomico di Brera, Via E. Bianchi 46 I-23807 Merate}
}
\author{G. Ghisellini}{
  address={Osservatorio Astronomico di Brera, Via E. Bianchi 46 I-23807 Merate}
}
\author{L. Nava}{
  address={Osservatorio Astronomico di Brera, Via E. Bianchi 46 I-23807 Merate}
}
\author{C. Firmani}{
  address={Osservatorio Astronomico di Brera, Via E. Bianchi 46 I-23807 Merate}
  ,altaddress={Instituto de Astronomia U.N.A.M., A.P. 70-264, 04510, Mexico, D.F., Mexico} 
}

\begin{abstract}
The use of Gamma Ray Bursts as ``standard candles'' has been made
possible by the recent discovery of a very tight correlation between
their rest frame intrinsic properties. This correlation relates the
GRB prompt emission peak spectral energy $E_{\rm peak}$ to the energy
$E_{\gamma}$ corrected for the collimation angle $\theta_{\rm jet}$ of
these sources. The possibility to use GRBs to constrain the
cosmological parameters and to study the nature of Dark Energy are
very promising.
\end{abstract}

\maketitle


\section{Introduction}

The extremely large luminosity of Gamma Ray Bursts (GRBs) makes them
detectable, in principle, out to very large redshifts $z<20$
(e.g. \cite{Lamb2000}).  The present redshift distribution for
$\sim$60 GRBs, which extends out to $z=$ 6.29 for GRB 050904
(\cite{Antonelli2005}) would make GRBs exquisite potential tools for
observational cosmology.  They can have a profound impact on: (i) the
study the epoch of re-ionization; (ii) the characterization of the
properties of the cosmic intervening medium; (iii) the study of the
cosmic star formation history back to unprecedented epochs; (iv) the
description of the geometry of the Universe and (v) the investigation
of the nature of Dark Energy (DE).
 
However, the last two points require a class of ``standard candles''
whose spread in the Hubble diagram is comparable (and even smaller)
than the precision on the measure of their luminosity distance. At
first glance GRBs are everything but standard candles: their intrinsic
isotropic emitted energies span more than 4 orders of magnitudes, and
even the collimation corrected energy span about two orders of
magnitudes.  This has prevented, until recently, their application as
cosmological tools (\cite{Bloom2001}; \cite{Frail2001};
\cite{Schaefer2003})

However, the discovery of a very tight (so called ``Ghirlanda'')
correlation, with a scatter less than 0.1 dex, between the rest frame
spectral peak energy $E_{\rm peak}$ and the collimation corrected
energy $E_{\gamma}$ (\cite{Ghirlanda2004}) allowed a very accurate
measurement of the true GRB energetics and made them usable as
``standard candles'' to constrain the cosmological parameters
$\Omega_{M},\Omega_{\Lambda}$ and to study the DE equation of state
(\cite{Ghirlanda2004a,Firmani2005}).
\begin{center}
\begin{figure}
\includegraphics[height=0.6\textheight]{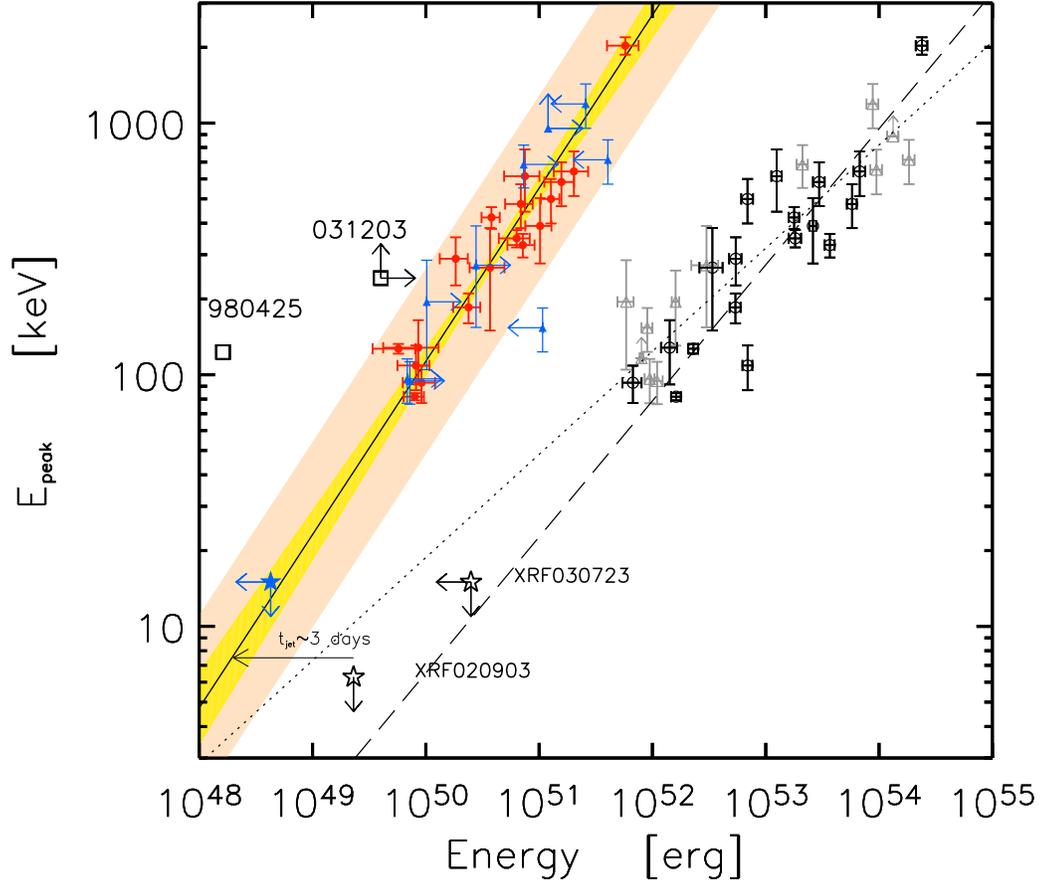}
\caption{ Rest frame peak energy $E_{\rm peak}$ versus isotropic (open
symbols) and collimation corrected (filled symbols) energy.  The black
open circles are the 18 GRBs (\cite{Nava2005}) with measured $z$,
$E_{\rm peak}$ and $t_{\rm jet}$ for which the collimation corrected
energy could be computed (red filled circles).  Upper/lower limits on
either one of the variables are shown by the blue filled triangles.
The best fit powerlaw to the red filled circles (Eq. 1) is represented
by the solid (blue) line and its uncertainty by the thin yellow shaded
region.  The large (light orange) shaded region represents the
$3\sigma$ gaussian scatter of the data points around the correlation.
Also shown is the Amati correlation obtained by fitting the open black
data points and the open grey triangles (which are not upper/lower
limits) either accounting for the errors on both the coordinates (long
dashed line - slope = 0.54 and $\chi^2_{\rm r}=8.4$ for 26 dof) and by
a linear regression (dotted line - slope = 0.4).  Note the two XRF
(030723 and 020903) and the two GRBs 980425/SN1998bw and
031203/SN2003lw which are outliers with respect to both the Ghirlanda
and the Amati correlations.  }
\label{fig_gg1}
\end{figure}
\end{center}
There are some issues to be discussed about the use of the Ghirlanda
correlation for cosmology and the existence of the $E_{\rm
peak}$-$E_{\gamma,\rm iso}$ correlation (so called ``Amati'' ref
\cite{Amati2002}).

The cosmological use of the $E_{\rm peak}$-$E_{\gamma}$ correlation
suffers from the so called ``circularity'' problem
(\cite{Ghirlanda2004}, see also \cite{Ghisellini2005}) due to the fact
that the correlation is not calibrated with (for instance) low
redshift GRBs and, therefore, its slope and normalization are
cosmology dependent.  The latter problem could be solved with either a
number of low redshift GRBs or through a convincing theoretical
interpretation (see \cite{Eichler2004,Levinson2005},
\cite{Yamazaki2004} and \cite{Rees2005}) for possible interpretations)
fixing at least its slope.  In the meanwhile different approaches have
been adopted (\cite{Ghirlanda2004a,Firmani2005}) to circumvent this
problem.

The Amati correlation was derived with a very limited sample of GRBs
(originally only 9 \cite{Amati2002}, then 24 \cite{Ghirlanda2004}) and
it is possible that it is affected by some selection effect connected
to the need to have a measured spectroscopic redshift (which may
select the brightest bursts).  While these possible selection effects
are still a matter of debate, this correlation (as well as the
Ghirlanda one) appears to be satisfied by the (few) newly discovered
Hete--II and Swift GRBs.  Furthermore, more sophisticated statistical
tests, based on the large BATSE sample of GRBs
(\cite{Nakar2005,Band2005} and \cite{Ghirlanda2005,Bosnjak2005}), have
been performed.

\section{The $E_{\rm peak}-E_{\gamma}$ correlation in long GRB}
The $E_{\gamma}$-$E_{\rm peak}$ correlation is found with all 
the GRBs with 
(i) a secure (spectroscopic) redshift measurement, 
(ii) well known prompt emission spectral properties and 
(iii) a measured jet break time (from the afterglow light curve).  
The latter, in fact,
allows to derive, within the standard GRB scenario, the burst opening
angle $\theta_{\rm jet}$ and to compute the collimation corrected
energy $E_{\gamma}=E_{\rm iso}(1-\cos\theta)$.

Since the publication of the initial sample of 15 GRBs
\cite{Ghirlanda2004} there have been 3 new GRBs with the
all the three input parameters measured. 
Furthermore, some spectral data have been slightly revised
for the old GRBs.
Then the sample now contains 
%
%
18 GRBs (updated to August 2005) with measured $z$,
$E_{\rm peak}$ and $t_{\rm jet}$. 
This sample is presented and discussed in details in \cite{Nava2005}.  
In the uniform jet model with a uniform density circumburst medium, 
the updated Ghirlanda correlation becomes:
\begin{equation}
\left({E^\prime_{\rm p} \over 100\, {\rm keV}}\right) \, =\,
(2.79\pm0.15)\,
\left({E_\gamma\over 2.72 \times 10^{50}\, {\rm erg}}\right)^{0.69\pm 0.04}
\label{ggl_new}
\end{equation}
with a reduced $\chi^2_{\rm r}=1.4$ (16 dof). This correlation and the
data points (solid line and red--filled circles, respectively) are
shown in Fig. 1 together with the updated Amati correlation (long
dashed line). 
The data points (red filled circles) have a gaussian
scatter with $\sigma=0.1$ (the 3$\sigma$ scatter is represented by the
light shaded region in Fig. 1) around the correlation represented by
Eq. 1. Both the scatter and the slope of this updated Ghirlanda
correlation are consistent with what found with the sample of 15
bursts of \cite{Ghirlanda2004}.

With this updated Ghirlanda correlation (Eq. 1) we fitted the
cosmological parameters $\Omega_{M},\Omega_{\Lambda}$ and
$w_{0},w_{a}$ with the bayesian method proposed in \cite{Firmani2005}
and found results fully consistent with those presented in \cite{Firmani2005}. 

\section{The jet opening angle distribution of GRBs}

\begin{center}
\begin{figure}
\includegraphics[height=0.5\textheight]{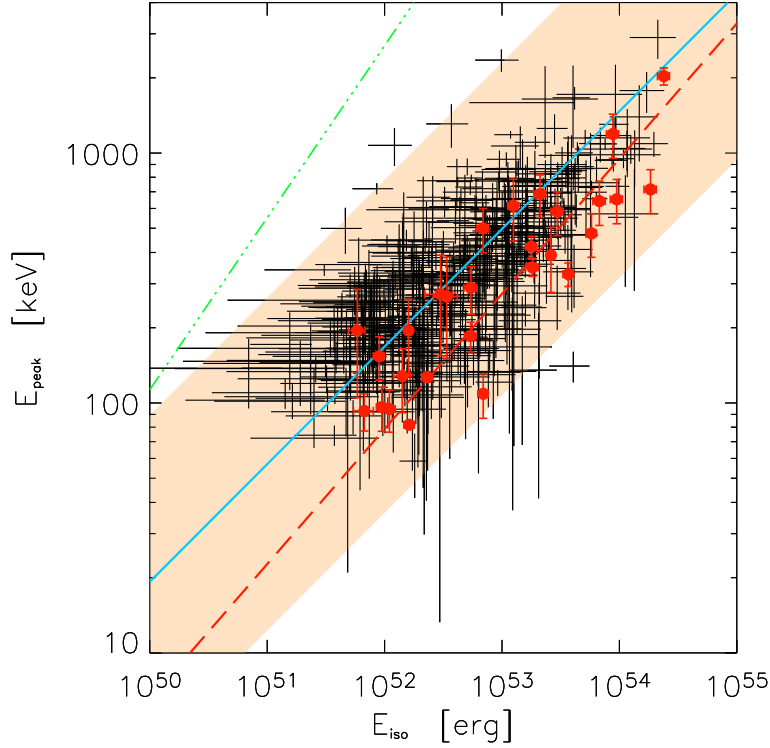} 
\caption{
Rest frame peak energy $E_{\rm peak}$ versus isotropic
equivalent energy $E_{\gamma, \rm iso}$. 
Red filled circles are the 28 GRBs with spectroscopically measured 
redshifts and published spectral properties. 
The long--dashed line is their best fit (weighting for the
errors on both variables) which is 
$E_{\rm peak}\propto E_{\gamma,\rm iso}^{0.54}$ 
with a reduced $\chi^2_{\rm r}=8.4$ (26 dof). 
The black crosses are the 442 GRBs with pseudo redshifts derived from the 
lag--luminosity relation. 
The solid line is the best fit to these data points which gives 
$E_{\rm peak}\propto E_{\gamma,\rm iso}^{0.47}$ with a reduced 
$\chi^2_{\rm r}=4.0$ (440 dof). 
The shaded area represents the 3$\sigma$ scatter region 
of the black points around their best fit line (solid line). 
The triple--dot--dashed line is the Ghirlanda correlation (Eq. 1).}
\end{figure}
\end{center}
The Ghirlanda correlation is derived by correcting the isotropic
equivalent energies $E_{\gamma,\rm iso}$ for the collimation angle
$\theta_{\rm jet}$. In \cite{Ghirlanda2004} (see also
\cite{Ghirlanda2005}) it was demonstrated that the scatter of the
Amati correlation can be interpreted as due entirely to the
different jet opening angles (see also Fig. 1 of \cite{Firmani2005}). 
Still, it has been argued
(\cite{Nakar2005,Band2005}) that the Amati correlation might be only a
selection effect because inconsistent with the largest population of
GRBs detected by BATSE. We performed (\cite{Ghirlanda2005} see also
\cite{Bosnjak2005}) a different test by checking if a large sample of
GRBs with a pseudo--redshift estimate  could indicate the existence of a
relation in the $E_{\rm peak}-E_{\gamma,\rm iso}$ plane. 
%
\begin{center}
\begin{figure}
  \includegraphics[height=0.5\textheight]{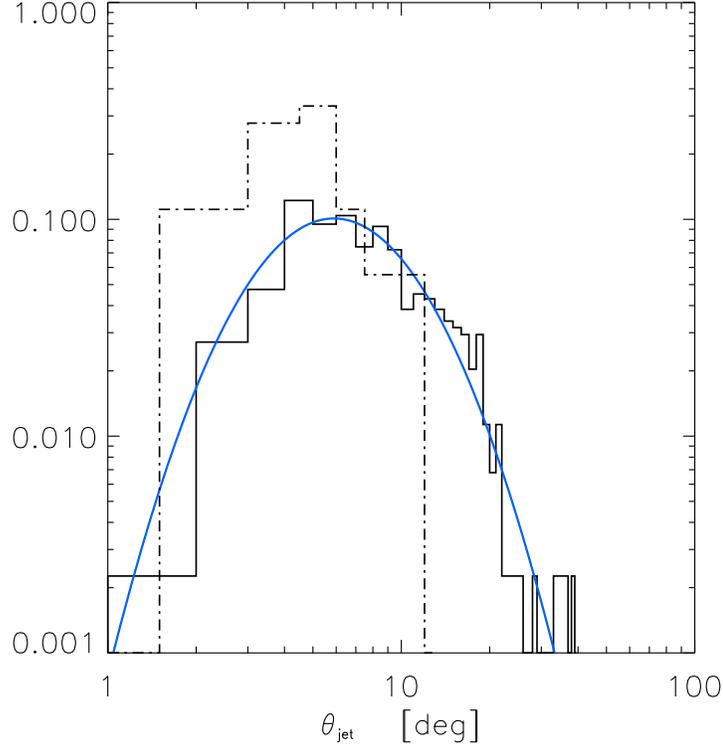}
\caption{
Jet opening angle distributions. The solid line histogram
  represents the $\theta_{\rm jet}$ derived from the large sample of
  442 GRBs with pseudo redshifts requiring that they satisfy the
  Ghirlanda relation as represented by Eq. 1. 
The solid line is the best fit log--normal distribution. 
The dot-dashed histogram represents the angle
  distribution of the 18 GRBs with spectroscopic redshifts and well
  constrained $t_{\rm jet}$. }
\end{figure}
\end{center}

With a sample of 442 long duration GRBs whose spectral properties has
been studied (\cite{Yonetoku2004}) and whose redshifts has been
derived from the lag--luminosity relation (\cite{Band2004}), we
populated the $E_{\rm peak}-E_{\gamma,\rm iso}$ plane (black crosses
in Fig. 2).  We found that this large sample of bursts produces an Amati
correlation (solid line in Fig. 2) whose slope (normalization) is
slightly flatter (larger) than that found with the sample of 28 GRBs
(red filled circles in Fig. 2). 
The gaussian scatter of the 442 bursts around their correlation has a standard
deviation $\sigma=0.22$, i.e. fully consistent with that of the 28 GRBs 
around their best fit correlation (long--dashed red line in Fig. 2). 
This suggests that a relation between $E_{\rm peak}$ and 
$E_{\gamma, \rm iso}$ does exist.

However, it might still be argued that the correlation found with the
28 and with the 442 GRBs have different normalizations, although
similar scatter and slopes. This is evident from the relative position
of the 28 GRBs (red filled circles in Fig. 2): they lie on the right
tail of the scatter distribution of the 442 GRBs (black crosses) in
the $E_{\rm peak}-E_{\gamma,\rm iso}$ plane. 
This can be easily interpreted as due to a selection effect.
In fact, we can derive the jet opening angle of the 442 GRBs 
by assuming that the Ghirlanda correlation exists and that its
scatter is (as shown in \cite{Ghirlanda2004,Firmani2005}) much smaller
than that of the Amati correlation.
The angle distribution is shown in Fig. 3 (solid histogram) and it is well
represented by a log--normal distribution (solid line) with a typical
$\theta_{\rm jet}\sim 6^{o}$. 
The angle distribution of the 28 GRBs
with measured $t_{\rm jet}$ is also shown (dot--dashed histogram in
Fig. 3) and it is shifted to the small--angle tail of the 
$\theta_{\rm jet}$ distribution of the other 442 bursts. 
This suggests that the 28 GRBs
which are used to define the Amati correlation have jet opening angles
which are systematically smaller than average.
This makes them more luminous and brighter.
In turn, this makes them better candidates to have an optical follow up
and to have their redshift measured.

We conclude that the origin of the Amati correlation is connected to:
i) the existence of the Ghirlanda correlation and
ii) the existence of a {\it peaked} jet opening angle 
distribution.
Were it flat, then the $E_{\rm \gamma, iso}-E_{\rm peak}$ correlation
would not exist.


\begin{theacknowledgments}
We thank the LOC and SOC for this stimulating conference.  We thank
A. Celotti, F. Tavecchio, D. Lazzati and D. Lamb for useful
discussions.
\end{theacknowledgments}

\end{document}